\documentclass[aps,preprint]{revtex4}%
\usepackage{amsfonts}
\usepackage{amsmath}
\usepackage{amssymb}
\usepackage{graphicx}%
\begin{document}
\title[ ]{Theory of Weiss oscillations in the magnetoplasmon spectrum of Dirac electrons
in graphene}
\preprint{ }
\author{M. Tahir$^{\ast}$}
\affiliation{Department of Physics, University of Sargodha, Sargodha, Pakistan}
\author{K. Sabeeh}
\affiliation{Department of Physics, Quaid-i-Azam University, Islamabad, Pakistan}
\author{}
\affiliation{}
\pacs{PACS number}

\begin{abstract}
We present the collective excitations spectrum (magnetoplasmon spectrum) of
Dirac electrons in a weakly modulated single graphene layer in the presence of
a uniform magnetic field. We consider electric modulation in one-dimension and
the magnetic field applied perpendicular to graphene.We derive analytical
results for the intra-Landau band plasmon spectrum within the
self-consistent-field approach. We find Weiss oscillations in the
magnetoplasmon spectrum which is the primary focus of this work. Results are
presented for the intra-Landau band magnetoplasmon spectrum as a function of
inverse magnetic field. These results are also compared with those of
conventional 2DEG. We have found that the Weiss oscillations in the
magnetoplasmon spectrum are larger in amplitude compared to those in
conventional 2DEG for the same modulation strength, period of modulation and
electron density.

\end{abstract}
\volumeyear{year}
\volumenumber{number}
\issuenumber{number}
\eid{identifier}
\date[Date text]{date}
\received[Received text]{date}

\revised[Revised text]{date}

\accepted[Accepted text]{date}

\published[Published text]{date}

\startpage{1}
\endpage{2}
\maketitle

\section{\textbf{INTRODUCTION }}

Recent progress in the experimental realization of a single layer of graphene
has lead to extensive exploration of electronic properties in this system.
Experimental and theoretical studies have shown that the nature of
quasiparticles in these two-dimensional systems is very different from those
of the conventional two-dimensional electron gas (2DEG) realized in
semiconductor heterostructures. Graphene has a honeycomb lattice of carbon
atoms. The quasiparticles in graphene have a band structure in which electron
and hole bands touch at two points in the Brillouin zone. At these Dirac
points the quasiparticles obey the massless Dirac equation. In other words,
they behave as massless Dirac fermions leading to a linear dispersion relation
$\epsilon_{k}=vk$ ( with the characteristic velocity $v\simeq10^{6}m/s)$. This
difference in the nature of the quasiparticles in graphene from conventional
2DEG has given rise to a host of new and unusual phenomena such as anamolous
quantum Hall effects and a $\pi$ Berry phase\cite{1}\cite{2}. These transport
experiments have shown results in agreement with the presence of Dirac
fermions. The 2D Dirac-like spectrum was confirmed recently by cyclotron
resonance measurements and also by angle resolved photoelectron spectroscopy
(ARPES) measurements in monolayer graphene\cite{3}. Recent theoretical work on
graphene multilayers has also shown the existance of Dirac electrons with a
linear energy spectrum in monolayer graphene\cite{4}.

It was found years ago that if conventional 2DEG is subjected to artificially
created periodic potentials in the submicrometer range it leads to the
appearence of Weiss oscillations in the magnetoresistance. This type of
electrical modulation of the 2D system can be carried out by depositing an
array of parallel metallic strips on the surface or through two interfering
laser beams \cite{5,6,7}. Weiss oscillations were found to be the result of
commensurability of the electron cyclotron diameter at the Fermi energy and
the period of the electric modulation. These oscillations were found to be
periodic in the inverse magnetic field \cite{5,6,7}. It is interesting to
study the affect of electrical modulation of graphene layer on the Dirac
electrons. Theoretical study of transport properties of Dirac electrons in
graphene was recently carried out which showed the appearance of enhanced
Weiss oscillations in the magnetoconductivity \cite{8}.

One of the important electronic properties of a system are the collective
excitations (plasmons). Weiss oscillations in the magnetoplasmon spectrum of
conventional 2DEG has been the subject of continued interest \cite{9}.
Plasmons in graphene were studied as early as in the eighties\cite{10} and
more recently \cite{11}. What distinguishes our work from earlier studies of
plasmons in graphene is that we consider the effect of electric modulation on
the magnetoplasmons which has not been studied before. In the present work, we
study the affect of electrical modulation of a graphene monolayer on the
magnetoplasmon spectrum within the self-consistent-field approach. We also
analyze the dynamic, nonlocal dielectric function of this system and in
particular consider the modulation-induced effects.

The present paper is arranged as follows. In section II we give the
formulation of the problem. We analytically determine the intra-Landau band
plasmon spectrum in section III. In section IV we present the numerical
results and discussion. Concluding remarks are made in section V.

\section{\ FORMULATION}

We consider two-dimensional Dirac electrons in graphene moving in the
x-y-plane. The magnetic field ($B_{z}$) is applied along the z-direction
perpendicular to the graphene plane. This system is subjected to weak electric
modulation along the x-direction. We employ the Landau gauge and write the
vector potential as $A=(0,Bx,0)$. The two-dimensional Dirac like Hamiltonian
for single electron in the Landau gauge is ($\hbar=c=1$ in this work)
\cite{1,2,8}%
\begin{equation}
H_{0}=v\sigma.(-i\nabla+eA) \label{1}%
\end{equation}
Here \ $\sigma=\{\sigma_{x},\sigma_{y}\}$are the Pauli matrices and $v$
characterizes the electron velocity. The complete Hamiltonian of our system is
represented as%
\begin{equation}
H=H_{0}+U(x) \label{2}%
\end{equation}
where $H_{0}$ is the unmodulated Hamiltonian and $U(x)$ is the one-dimensional
periodic modulation potential along the x-direction modelled as%
\begin{equation}
U(x)=V_{0}Cos(Kx) \label{3}%
\end{equation}
here $K=2\pi/a,a$ is the period of modulation and $V_{0}$ is the constant
modulation amplitude. The Landau level energy eigenvalues without modulation
are given by%
\begin{equation}
\varepsilon(n)=v\sqrt{2eB\left\vert n\right\vert }=\omega_{g}\sqrt{\left\vert
n\right\vert } \label{4}%
\end{equation}
where $\omega_{g}=v\sqrt{2eB}$is the cyclotron frequency of the graphene
electrons and $n$ is an integer$.$The Landau level spectrum for Dirac
electrons is significantly different from the spectrum for electrons in
conventional 2DEG\ which is given as $\varepsilon(n)=\omega_{c}(n+1/2)$, where
$\omega_{c}$ is the cyclotron frequency.

Since we are considering weak modulation such that $V_{0}$ is about an order
of magnitude smaller than the Fermi energy $E_{F}$, we can apply standard
perturbation theory to determine the first order correction to the unmodulated
energy eigenvalues in the presence of modulation with the result%
\begin{equation}
U_{n}=\frac{1}{2}V_{0}Cos(Kx_{0})e^{-u/2}[L_{n}(u)+L_{n-1}(u)] \label{5}%
\end{equation}
where $u=K^{2}l^{2}/2,$ $x_{0}=l^{2}k_{y}$, $L_{n}(u)$ are Laguerre
polynomials and $l=\sqrt{1/eB}$ is the magnetic length. Hence the energy
eigenvalues in the presence of modulation are
\begin{equation}
\varepsilon(n,x_{0})=\varepsilon(n)+U_{n}=v\sqrt{2eB\left\vert n\right\vert
}+F_{n}Cos(Kx_{0}) \label{6}%
\end{equation}
with\ $F_{n}=\frac{1}{2}V_{0}e^{-u/2}[L_{n}(u)+L_{n-1}(u)]$. This result has
been obtained previously \cite{8}. The degeneracy of the Landau level spectrum
of the unmodulated system with respect to $k_{y}$ is lifted in the presence of
modulation. The formerly sharp Landau levels broaden into bands. The Landau
bandwidths $\sim$ $F_{n}$ oscillates as a function of $n$ since $L_{n}(u)$ are
oscillatory functions of the index $n$. Similar features were noted in the
Landau level spectrum of conventional electrons in density modulated 2DEG. The
difference in the case of density modulated Dirac electrons is firstly that
the Landau level spectrum depends on the square root of both the magnetic
field $B$ and the Landau band index $n$ as against linear dependence in case
of conventional electrons. Secondly, for Dirac electrons the energy
eigenvalues in the presence of modulation given by Eq. (6) contain a term
which is a linear combination of Laguerre polynomials with indices $n$ and
$n-1$ while for conventional electrons there is only a single term that
contains Laguerre polynomial with index $n$. We expect that this modulation
induced change in the electronic density of states to influence the dielectric
response and the collective excitations of the graphene system.

\section{INTRA-LANDAU-BAND PLASMON\ SPECTRUM OF GRAPHENE}

The dynamic and static response properties of an electron system are all
embodied in the structure of the density-density correlation function. We
employ the Ehrenreich-Cohen self-consistent-field (SCF) approach\cite{12} to
calculate the density-density correlation function. The SCF treatment
presented here is by its nature a high density approximation which has been
successfully employed in the study of collective excitations in
low-dimensional systems both with and without an applied magnetic field. It
has been found that SCF predictions of plasmon spectra are in excellent
agreement with experimental results. Following the SCF approach, one can
obtain the dielectric function as%
\begin{equation}
\epsilon(\bar{q},\omega)=1-v_{c}(\bar{q})\Pi(\bar{q},\omega). \label{7}%
\end{equation}
where $v_{c}(\bar{q})=\frac{2\pi e^{2}}{\kappa\overline{q}}$, $\overline
{q}=(q_{x}^{2}+q_{y}^{2})^{1/2},\kappa$ is the background dielectric constant
and $\Pi(\bar{q},\omega)$ is the non-interacting density-density correlation
function
\begin{align}
\Pi(\bar{q},\omega)  &  =\frac{1}{\pi al^{2}}\sum C_{nn^{\prime}}(\frac
{\bar{q}^{2}}{2eB})\int\limits_{0}^{a}dx_{0}[f(\varepsilon(n,x_{0}%
+x_{0}^{\prime})-f(\varepsilon(n^{\prime},x_{0}))]\nonumber\\
&  \times\lbrack\varepsilon(n,x_{0}+x_{0}^{\prime})-\varepsilon(n^{\prime
},x_{0})+\omega+i\eta]^{-1}, \label{8}%
\end{align}
where $x_{0}=l^{2}k_{y}$, $x_{0}^{\prime}=l^{2}q_{y}$, $C_{nn^{\prime}}%
(\frac{\bar{q}^{2}}{2eB})=(n_{2}!/n_{1}!)\left(  \frac{\bar{q}^{2}}%
{2eB}\right)  ^{n_{1}-n_{2}}\left[  L_{n_{2}}^{^{n_{1}-n_{2}}}\right]  ^{2}$
with $n_{1}=\max(n,n^{\prime}),n_{2}=\min(n,n^{\prime})$ and $L_{n}^{^{l}}(x)$
an associated Laguerre polynomial. In writing the above equation we converted
the $k_{y}$-sum into an integral over $x_{0}$ by the following relation:
$\frac{1}{A}\underset{k_{y}}{\sum}\cdots\rightarrow2\frac{1}{2\pi al^{2}}%
\int\limits_{0}^{a}dx_{0}\cdots.$ This is a convenient form of $\Pi(\bar
{q},\omega)$ that facilitates writing of the real and imaginary parts of the
correlation function.

The real and imaginary parts of the density-density correlation function are
the essential ingredients in the theoretical considerations of such diverse
problems as high frequency and steady state transport, static and dynamic
screening and correlation phenomena. The plasmon modes are determined from the
roots of the longitudinal dispersion relation
\begin{equation}
1-v_{c}(\bar{q})\operatorname{Re}\Pi(\bar{q},\omega)=0, \label{9}%
\end{equation}
along with the condition Im$\Pi(\bar{q},\omega)=0$ to ensure long-lived
excitations. Employing Eq.(8), Eq.(9) can be expressed as
\begin{equation}
1=\frac{2\pi e^{2}}{k\bar{q}}\frac{1}{\pi al^{2}}\underset{n,n^{\prime}}{\sum
}C_{nn^{\prime}}(\frac{\bar{q}^{2}}{2eB})(I_{1}(\omega)+I_{1}(-\omega)),
\label{10}%
\end{equation}
with
\begin{equation}
I_{1}(\omega)=P\int\limits_{0}^{a}dx_{0}\frac{f(\varepsilon(n,x_{0}%
))}{\varepsilon(n,x_{0})-\varepsilon(n,x_{0}+x_{0}^{\prime})+\omega}
\label{11}%
\end{equation}
($P$ denotes principal value integral). The plasmon modes originate from two
kinds of electronic transitions: those involving different Landau bands
(inter-Landau band plasmons) and those within a single Landau-band
(intra-Landau band plasmons).

Inter-Landau band plasmons involve the local 1D magnetoplasma mode and the
Bernstein-like plasma resonances [9], all of which involve excitation
frequencies greater than the Landau-band separation. On the other hand,
intra-Landau band plasmons resonate at frequencies comparable to the
bandwidths, and the existence of these modes is due to finite width of the
Landau levels. The occurrence of such intra-Landau band plasmons is
accompanied by SdH type of oscillatory behavior in $1/B$. These SdH
oscillations occur as a result of emptying out of electrons from successive
Landau bands when they pass through the Fermi level as the magnetic field is
increased. The amplitude of these oscillations is a monotonic function of the
magnetic field, when the Landau bandwidth is independent of the band index
$n.$ In the density modulated case, the Landau bandwidths oscillate as a
function of the band index $n,$ so it is fair to expect that such oscillating
bandwidths would affect the plasmon spectrum of the intra-Landau band type,
resulting in new kind of oscillations called Weiss oscillations studied
extensively in conventional 2DEG. Here, we study these Weiss oscillations in
the magnetoplasmon spectrum in graphene.

To examine the intra-Landau-band plasmon spectrum we consider electronic
transitions within a single Landau band. In this case $V_{n}\ll\varepsilon
_{n}$, this approximation enables us to expand the argument of the
distribution function about $\varepsilon_{n}$ and substituting this expansion
into Eq.(11) yields
\begin{align}
I_{1}(\omega)  &  =\frac{1}{\omega}\int\limits_{0}^{a}dx_{0}f(\varepsilon
_{n}+F_{n}\cos(Kx_{0}))\nonumber\\
&  \times(1-\frac{F_{n}}{\omega}\{\cos(Kx_{0})-\cos(K(x_{0}+x_{0}^{\prime
}))\}). \label{12}%
\end{align}
The first integral in the above equation vanishes while we fold the range of
integration from 0 to $a/2$ by multiplying a factor of 2 in the second
integral. Hence we can express the intra-Landau band plasmon dispersion
relation, from Eqs. (10, 11, 12) and C$_{nn^{\prime}}(x)\rightarrow1$ for
$n=n^{\prime},$ as%
\begin{align}
1  &  =-\frac{16e^{2}}{k\bar{q}}\frac{1}{\pi al^{2}\omega^{2}}\sin^{2}%
(\frac{\pi}{a}x_{0}^{\prime})\nonumber\\
&  \times\underset{n}{\sum}F_{n}\int\limits_{0}^{a/2}dx_{0}f(\varepsilon
_{n}+F_{n}\cos(Kx_{0}))\cos(Kx_{0}) \label{13}%
\end{align}
This can be written as
\begin{equation}
\omega^{2}=\frac{16e^{2}}{k\bar{q}}\frac{1}{\pi al^{2}}\sin^{2}(\frac{\pi}%
{a}x_{0}^{\prime})\times\underset{n}{\sum\mid}F_{n}\mid A_{n}, \label{14}%
\end{equation}
here $A_{n}=\frac{a}{2\pi}\frac{\left\vert F_{n}\right\vert }{F_{n}}%
\sqrt{1-\Delta_{n}^{2}\text{ }}\theta(1-\Delta_{n})$ which results from the
analytic evaluation of the Fermi function integral in Eq.(13) that can be
carried out at $T=0.$The integral contributes only for $\Delta_{n}<1,$ where
$\Delta_{n}$ is given by $\Delta_{n}=\left\vert \frac{\varepsilon
_{F}-\varepsilon_{n}}{F_{n}}\right\vert $ with $\varepsilon_{F}$ the Fermi
energy. Hence the zero temperature intra-Landau-band plasmon dispersion
relation reduces to
\begin{equation}
\omega^{2}=\frac{8e^{2}}{k\bar{q}}\frac{1}{\pi al^{2}}\sin^{2}(\frac{\pi}%
{a}x_{0}^{\prime})\times\underset{n}{\sum}\left\vert F_{n}\right\vert
\sqrt{1-\Delta_{n}^{2}}\theta(1-\Delta_{n}). \label{15}%
\end{equation}
In general, the inter- and intra-Landau-band modes are coupled for arbitrary
magnetic field strength. So long as $\left\vert F_{n}\right\vert <v\sqrt
{neB},$ mixing of the inter-and intra- Landau band modes is small. Only the
intra-Landau-band mode ($\omega)$ will be excited in the frequency regime
$v\sqrt{neB}>\omega\sim$ $\left\vert F_{n}\right\vert .$

\section{NUMERICAL RESULTS AND DISCUSSION}

The intra-Landau-band plasma frequency for Dirac electrons given by equation
(15) is shown graphically (solid curve) in Fig.1 as a function of $1/B$. In
the same figure we also display the intra-Landau-band plasma frequency for
standard electrons in conventional 2DEG (dashed curve). The following
parameters were employed for graphene: $k=2.5$, $n_{D}=3\times10^{15}$
m$^{-2}$, $v=2.6$ eV \AA , $a=382$ nm and $V_{0}=0.5$ meV. We also take
$q_{x}=0$ and $q_{y}=.01k_{F},$ with $k_{F}=(\pi n_{D})^{1/2}$.$\ $For
conventional 2DEG (a 2DEG at the GaAs-AlGaAs heterojunction) we use the
following parameters: $m=.07m_{e}$($m_{e}$ is the electron mass), $k=12$ and
$n_{D}=3\times10^{15}$ m$^{-2}$ with the modulation strength and period same
as in the graphene system. In Fig.1, the modulation induced oscillations due
to the 1D modulation are clearly visible superposed on the sharp SdH-type
oscillations. The intra-Landau-band plasmons have frequencies comparable to
the bandwidth and they occur as a result of broadening of the Landau levels
due to the modulation in our system. These type of intra-Landau-band plasmons
accompanied by regular oscillatory behavior (in $1/B)$ of the SdH type was
first predicted in \cite{13} for tunneling planar superlattice where the
overlap of electron wavefunction in adjacent quantum wells provides the
mechanism for broadening of Landau levels. The SdH oscillations occur as a
result of emptying out of electrons from successive Landau levels when they
pass through the Fermi level as the magnetic field is increased. The amplitude
of these oscillations is a monotonic function of the magnetic field when the
Landau bandwidth is independent of the band index $n.$ In the density
modulated case, the Landau bandwidths oscillate as a function of the band
index $n,$ with the result that in the plasmon spectrum of the intra-Landau
band type, there is a new kind of oscillation called Weiss oscillation which
is also periodic in $1/B$ but with a different period and amplitude from the
SdH type oscillation.

The origin of these two types of oscillations can also be understood by a
closer analytic examination of equation (15). In the regime, $v\sqrt{neB}>$
$\left\vert F_{n}\right\vert $, the unit step function vanishes for all but
the highest occupied Landau band, corresponding, say, to the band index $N$.
The sum over $n$ is trivial, and plasma frequency is given as $\omega
\sim\left\vert F_{N}\right\vert ^{1/2}(1-\Delta_{N}^{2})^{1/4}\theta
(1-\Delta_{N})$. The analytic structure primarily responsible for the SdH
oscillations is the function $\theta(1-\Delta_{N})$, which jumps periodically
from zero (when the Fermi level is above the highest occupied Landau band) to
unity (when the Fermi level is contained with in the highest occupied Landau
band). On the other hand, Weiss oscillations which represent the periodic
modulation of the amplitude of SdH oscillation are largely a consequence of
the oscillatory factor $\left\vert F_{N}\right\vert ^{1/2}$.

From the results displayed in Fig.1, we can also make a comparison of the
Weiss oscillations in the magnetoplasmon spectrum of Dirac electrons in
graphene and standard electrons in conventional 2DEG. For the same modulation
strength, period of modulation and electron density we find that the Weiss
oscillations in graphene have a larger amplitude compared to conventional
2DEG. This can be attributed to the larger characteristic velocity
($v\sim10^{6}m/s)$ of electrons in graphene compared to the Fermi velocity of
standard electrons and smaller background dielectric constant $k$ in graphene
in contrast to conventional 2DEG.

The SCF theory presented here is expected to be a good quantitative
approximation since the dimensionless electron density parameter $r_{s}$ in
graphene is a constant ($<1)$\cite{11}, hence we are in the high density
regime. It should also be noted that for electron densities under
consideration the effective Fermi temperature $T_{F}=E_{F}/k_{B}$ in graphene
is very high ($\sim1300K$ for $n_{D}\sim10^{16}$ m$^{-2})$, so our $T=0$
theory is expected to be valid at higher temperatures. The only constraint is
that the intra-Landau band plasmons are due to electronic transitions within a
single Landau band and the existence of these modes is due to finite width of
the Landau levels caused by the modulation. Therefore modulation induced
effects will be observable for temperatures where the thermal energy is less
than the Landau band width such that $k_{B}T<2|F_{n}|.$ In effect the theory
is valid for experimental situations where $k_{B}T$ is smaller than
$\omega_{g}$, $F_{n}$ and $\varepsilon_{F}.$

\section{CONCLUSIONS}

We have determined the intra-Landau band plasma frequency for Dirac electrons
in a modulated graphene layer in the presence of a magnetic field employing
the SCF approach. We show that the intra-Landau band magnetoplasmon spectrum
exhibits the electronic density of states modulation in the form of
oscillating magnetoplasmon frequencies (Weiss oscillations). We also compare
these results with those of the conventional 2DEG. We have found that the
Weiss oscillations in the magnetoplasmon spectrum are larger in amplitude
compared to those in conventional 2DEG for the same modulation strength,
period of modulation and electron density.

\section{Acknowledgements}

One of us (K.S.) would like to acknowledge the support of the Pakistan Science
Foundation (PSF) through project No. C-QU/Phys (129). M. T. would like to
acknowledge the support of the Pakistan Higher Education Commission (HEC).

$\ast$Present address: Department of Physics, Blackett Laboratory, Imperial
College London, London SW7 2AZ, United Kingdom.

\end{document}